\documentclass[intlimits,twoside,a4paper]{article}
\usepackage[cp1251]{inputenc}
\usepackage[eqsecnum]{cmpj3}

\issue{2022}{25}{4}{43703}
\doinumber{10.5488/CMP.25.43703}

\title[Critical exponents of the order parameter of diffuse ferroelectric phase transitions]{Critical exponents of the order parameter of diffuse ferroelectric phase transitions in the solid solutions based on lead germanate: studies of optical rotation}
\author[D. I. Adamenko, R. O. Vlokh]{D. I. Adamenko\orcid{0000-0002-0541-8233}\thanks{\email{grdmagn2017@gmail.com}.}, R. O. Vlokh\orcid{0000-0001-6048-6370}}
\address{Vlokh Institute of Physical Optics, 23 Dragomanov St., 79005 Lviv, Ukraine}


\date{Received June 08, 2022, in final form August 17, 2022}

\begin{document}

\maketitle

\begin{abstract}
In this work we show that the critical exponents of the order parameter (CEOPs) of diffuse ferroelectric phase transitions (DFEPTs) occurring in lead germanate-based crystals can be determined using experimental temperature dependences of their optical rotation. We also describe the approach that suggests dividing a crystal sample into many homogeneous unit cells, each of which is characterized by a non-diffuse phase transition with a specific local Curie temperature. Using this approach, the CEOPs have been determined for the pure Pb\textsubscript{5}Ge\textsubscript{3}O\textsubscript{11} crystals, the solid solutions Pb\textsubscript{5}(Ge$_{1-x}$Si$_{x}$)\textsubscript{3}O\textsubscript{11} ($x = $~0.03, 0.05, 0.10, 0.20, 0.40) and (Pb$_{1-x}$Ba$_{x}$)\textsubscript{5}Ge\textsubscript{3}O\textsubscript{11} ($x =$ 0.02, 0.05), and the doped crystals Pb\textsubscript{5}Ge\textsubscript{3}O\textsubscript{11}:Li\textsuperscript{3+} (0.005 wt.~\%),  Pb\textsubscript{5}Ge\textsubscript{3}O\textsubscript{11}:La\textsuperscript{3+} (0.02 wt.~\%),  Pb\textsubscript{5}Ge\textsubscript{3}O\textsubscript{11}:Eu\textsuperscript{3+} (0.021~wt.~\%),  Pb\textsubscript{5}Ge\textsubscript{3}O\textsubscript{11}:Li\textsuperscript{3+},  Bi\textsuperscript{3+} (0.152 wt.~\%) and Pb\textsubscript{5}Ge\textsubscript{3}O\textsubscript{11}:Cu\textsuperscript{2+} (0.14 wt.~\%). Comparison of our approach with the other techniques used for determining the Curie temperatures and the CEOPs of DFEPTs testifies to its essential advantages.
\printkeywords
\end{abstract}

\section{Introduction}

Lead germanate crystals Pb\textsubscript{5}Ge\textsubscript{3}O\textsubscript{11} (abbreviated hereafter as PGO) exhibit a proper second-order ferroelectric phase transition (PT) at the Curie temperature $T_{\textrm{C}} \approx 450$~K~\cite{Iwas72}. At $T > T_{\textrm{C}}$, the crystals belong to a hexagonal system (the point symmetry group $\bar{6}$). The sixth-fold inversion symmetry axis vanishes at $T < T_{\textrm{C}}$ and the symmetry becomes trigonal (the point group 3).

Ferroelectric properties of PGO were discovered nearly 50 years ago. In spite of a long history of their studies, the crystals still attract a considerable attention of researchers~\cite{Vazh98,Novi00,Trub00,Roli01,Adam08}. Probably, this is partially due to the fact that PGO remains a unique example of materials where the PT is accompanied by the symmetry change $\bar{6} \leftrightarrow 3$. Moreover, this symmetry change is very convenient while studying the optical rotatory power. Indeed, the optical rotation in PGO can be directly measured for the light with wavelength~$\lambda$ propagating along the optic axis, with no accompanying linear optical birefringence. In addition, the PGO crystals represent a basis for a large family of solid solutions and doped crystals~\cite{Vlkh84}, whereas replacement of chemical elements in PGO or its doping can be achieved using fairly simple technological processes. At the same time, such solid solutions and doped crystals are attractive objects for the study of diffuse ferroelectric PTs (DFEPTs) --- namely, ferroelectric PTs that do not have a point character, but occur in certain, more or less expressed, temperature intervals (diffusion regions)~\cite{Rolv83}.

Below the Curie temperature, PGO becomes optically active due to the effect of electrogyration induced by spontaneous electric polarization~\cite{Vlkh84}:
\begin{equation}\label{g33}
g_{33} = \gamma_{333} (P_{\textrm{S}})_{3},
\end{equation}
where $g_{33}$ is the gyration-tensor component, $\gamma_{333}$ is the spontaneous electrogyration coefficient and $(P_{\textrm{S}})_{3}$ is the spontaneous polarization. The spontaneous polarization $(P_{\textrm{S}})_{3} \equiv P_{\textrm{S}}$ represents the order parameter of the PT. It is linearly related to the optical rotation $\rho_{3} \equiv \rho$, which can be defined as a specific rotation angle of the polarization plane of light~\cite{Iwas72,Vazh98,Novi00,Trub00,Roli01,Adam08,Vlkh84,Aizu64}. Therefore, the studies of temperature dependence of the optical rotation, which can be considered as a spontaneous electrogyration, enable one to derive many characteristics of the PT.

Despite diverse knowledge about the physical properties of PGO, some aspects of their critical behavior are still unclear. First of all, this applies to experimental determination of critical exponent of the order parameter (CEOP) $\beta$. This parameter can be found from the temperature dependence of spontaneous polarization:
\begin{equation}\label{Ps}
P_{\textrm{S}} \propto (T_{\textrm{C}} - T)^{\beta}.
\end{equation}
In the framework of mean-field Landau theory for the proper second-order ferroelectric PTs, $\beta$ should be equal to 0.5.

In 1972, Iwasaki et al.~\cite{Iwas72} analyzed the experimental dependence $P_{\textrm{S}} = P_{\textrm{S}}(T)$ for the PGO crystals and found that its behavior corresponds to the classical Landau theory [i.e., $P_{\textrm{S}} \propto (T_{\textrm{C}} - T)^{0.5}$] only in the region $T_{\textrm{C}} - T < 30$~K below the point $T_{\textrm{C}} = 450$~K. Further on, Konak et al.~\cite{Konk78} noted that the temperature dependence of the optical rotation $\rho$ in PGO is described by the ``empirical'' relation $\rho \propto (T_{\textrm{C}} - T)^{0.35}$ in the region $T_{\textrm{C}} - T > 3$~K (with $T_{\textrm{C}} = 450$~K). In other words, although $\rho$ represents a secondary order parameter of the PT, its behavior differs significantly from that predicted by the Landau theory. In 1999, Trubitsyn et al.~\cite{Trub99} showed that the splitting parameter $\Delta B$ of one of the spectral EPR lines of probing Gd\textsuperscript{3+} ions in PGO behaves as a ``local'' order parameter of the PT according to the Landau theory [$\Delta B \propto (T_{\textrm{C}} - T)^{0.5}$] in a sufficiently wide temperature region ($T_{\textrm{C}} - T < 150$~K) below the Curie temperature ($T_{\textrm{C}} = 451.4$~K).

In 2005, Shaldin et al.~\cite{Shal05} studied the temperature behavior of spontaneous polarization for PGO in the temperature range from 4.2 to 300~K. Their experimental results and the literature data available at the time allowed the authors of reference~\cite{Shal05} to detect the changes in the critical behavior of PGO from a dipole type ($\beta = 0.5$) to a pseudo-quadrupole type ($\beta = 0.25$) with increasing temperature from 290 to $T_{\textrm{C}} = 450$~K. This manifests itself as a change in the behavior of the parameter $P_{\textrm{S}}^{1/\beta}$ as a function of $(T_{\textrm{C}} - T)$ at $T_{\textrm{C}} - T = 50$~K. In 2006, Miga et al.~\cite{Miga06} obtained the CEOP $\beta = 0.51 ± 0.03$ and the Curie temperature $T_{\textrm{C}} = 452.58 ± 0.03$~K, using the experimental temperature dependence of the residual polarization $P_{\textrm{R}}$ and its fitting by the formula $P_{\textrm{R}} \propto (T_{\textrm{C}} - T)^{\beta}$ in the temperature region $T_{\textrm{C}} - T < 12$~K. These parameters were obtained after a sample under study was aged in the electric field with the strength $10^{6}$~V/m. Finally, in 2008, Kushnir et al.~\cite{Kush08} performed the optical studies of fluctuations of the order parameter for Pb\textsubscript{5}(Ge$_{1-x}$Si$_{x}$)\textsubscript{3}O\textsubscript{11} (PGSO) and (Pb$_{1-x}$Ba$_{x}$)\textsubscript{5}Ge\textsubscript{3}O\textsubscript{11} (PBaGO) solid solutions. They demonstrated a need in excluding the region of strong fluctuations from the analysis when finding consistent values of the CEOPs $\beta$. The region where it is advisable to perform $\beta$ calculations was estimated as $G \leqslant (T_{\textrm{C}} - T) / T_{\textrm{C}} \leqslant G^{1/3}$, where $G$ is Ginzburg's number. For the pure PGO crystals $T_{\textrm{C}} = 447$~K and $G = 0.01$, given that $4.5 \leqslant (T_{\textrm{C}} - T) \leqslant 96.3$~K. Then, the $\beta$ value can be determined as a slope of linear part of the dependence
\begin{equation}\label{beta}
\log \rho \propto \beta \log (T_{\textrm{C}} - T),
\end{equation}
which is equal to 0.44 for pure PGO. Moreover, the $\beta$ values for the PGSO and PBaGO solid solutions, which manifest diffuse ferroelectric phase transitions (DFEPT), deviate even more significantly from the classical value $\beta = 0.5$.

Considering the methods available for determining CEOPs of the PTs, we limit ourselves to the methods that analyze the temperature behavior of the optical rotation $\rho$ and use the dependences like equation~\eqref{beta} to calculate the CEOP $\beta$. This is because in reality the other approaches determine only the temperature region where the spontaneous electric polarization or some secondary order parameter of the PT is proportional to $(T_{\textrm{C}} - T)^{0.5}$. As can be seen from equation~\eqref{beta}, exact setting of the Curie temperature $T_{\textrm{C}}$ is essential for such methods.

Within the framework of optical techniques, the Curie temperature is usually chosen as a point where the optical rotation $\rho$ disappears completely in the process of its temperature change. Hereinafter, this approach is referred to as a method I. On the other hand, in reference~\cite{Kush08} the $T_{\textrm{C}}$ parameter is defined as a point of minimum of the temperature dependence of the derivative $\rd \rho^{2} / \rd T$. This approach is called a method~II.

When the ferroelectric PT is diffuse, the Curie temperature cannot be defined unambiguously. This can lead to significant errors in determining the CEOP $\beta$ even though one excludes a temperature region near the PT from the calculations described by equation~\eqref{beta}.

\begin{figure}[htb]
\centerline{\includegraphics[width=0.45\textwidth]{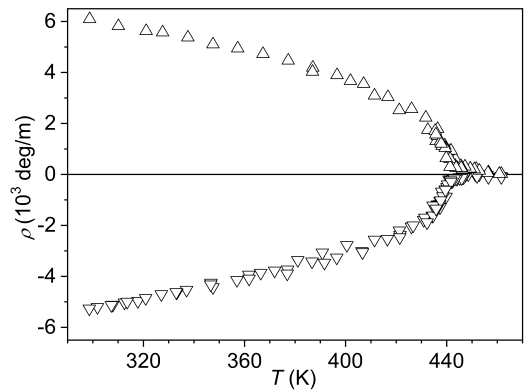}}
\caption{Temperature dependences of optical rotation $\rho$ for enantiomorphous domains of the doped PGO:Cu\_140 crystals ($\lambda = 632.8$~nm)~\cite{Adam09}.}
\label{domnCu}
\end{figure}

\begin{figure}[htb]
\centerline{\includegraphics[width=0.45\textwidth]{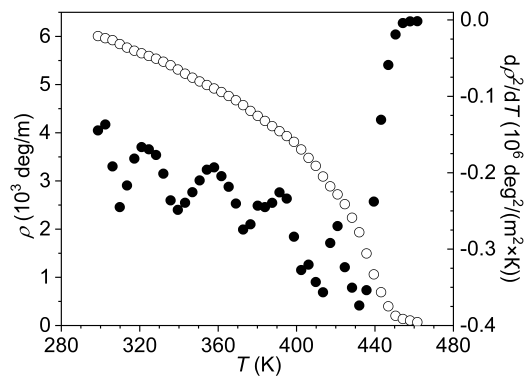}}
\caption{Smoothed temperature dependences of optical rotation $\rho$ (open circles) and its derivative $\rd \rho^{2} / \rd T$ (full circles) for the doped PGO:Cu\_140 crystals ($\lambda = 632.8$~nm).}
\label{smthCu}
\end{figure}

\begin{figure}[htb]
\centerline{\includegraphics[width=0.45\textwidth]{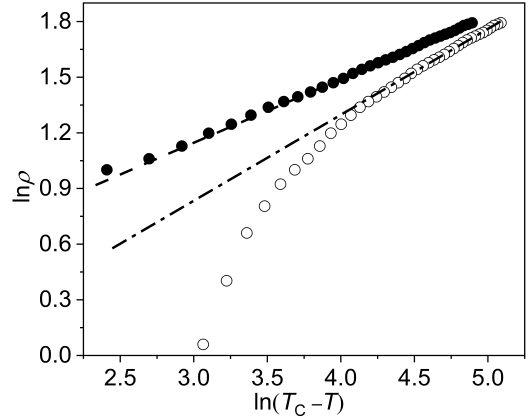}}
\caption{Dependences $\ln \rho$ vs. $\ln (T_{\textrm{C}} - T)$ for different methods of choosing the $T_{\textrm{C}}$ point (see also the text) in doped PGO:Cu\_140 crystals ($\lambda = 632.8$~nm).}
\label{betaCu}
\end{figure}

We illustrate the ambiguity in the choice of the Curie temperature based on the example of DFEPT in PGO doped with 0.140 wt.~\% of Cu\textsuperscript{2+} ions (PGO:Cu\_140), which was studied in our recent work~\cite{Adam09}. As seen from figure~\ref{domnCu}, the temperature dependences of the optical rotation $\rho$ in the PGO:Cu\_140 crystals have a sufficiently long ``tail'', which indicates to the presence of DFEPT. Using the method I, $T_{\textrm{C}}(\textrm{I}) = 460.8$~K can be found (figure~\ref{smthCu}) and the CEOP $\beta(\textrm{I})$ determined with equation~\eqref{beta} is equal to 0.47 (see open circles and dash-dotted line in figure~\ref{betaCu}). On the other hand, the method II results in $T_{\textrm{C}}(\textrm{II}) = 432.0$~K (figure~\ref{smthCu}) and $\beta(\textrm{II}) = 0.34$ (see full circles and dashed line in figure~\ref{betaCu}). Of course, both calculations are based on the temperature regions where the dependence $\ln \rho$ vs. $\ln (T_{\textrm{C}} - T)$ is linear. Moreover, the CEOPs $\beta$ found after choosing the Curie temperature $T_{\textrm{C}}$ by the methods I and II differ significantly.

The circumstances mentioned above reveal a need in finding some other approaches to determine the DFEPT parameters for the solid solutions and doped crystals of PGO family. This point is the main purpose of this work.

\section{Phenomenological approach}

The technique used in this work is based on the following: (i) a generalized model of diffuse PTs, according to which a crystal under study can be divided into infinitely large number of homogeneous unit cells so that the PT in each of these cells is not diffuse and manifests a specific local Curie temperature~\cite{Rolv83}; (ii)~Gaussian distribution of the local Curie temperatures in the homogeneous unit cells due to the central limit theorem, where the role of mathematical expectation is played by the so-called average Curie temperature $\Theta$, which is taken as the PT point and characterizes the state when half of the sample undergoes the PT~\cite{Smol85}; (iii)~a general relation~\eqref{Ps} for the order parameter of a proper second-order ferroelectric PT and its CEOP. Assume that diffusion of the PT is caused by some scalar inhomogeneity, e.g., scalar defects that do not change the symmetry of crystal matrix and affect only the Curie temperature distribution in the sample.

Although there are a number of works considering the local properties that affect the critical behaviour (see e.g.,~\cite{Grin76}) the approach consisting in dividing the crystals into non-interacting homogeneous cells is proper for the works mentioned above~\cite{Rolv83,Smol85}. Of course, even Ising's model is a particular case of work~\cite{Grin76}, which can be considered as the dividing of the system into the subsystems, while these subsystems interact between each other.

Let the diffusion region of the phase transition $\Delta T$ contain  $N \in \{2,~3,~4,\ldots{}\}$ local Curie temperatures $T_{\textrm{C}i}$ ($i = 1, \ldots{}, N)$. Then, we have
\begin{equation}\label{Temp}
\Delta T = T_{\textrm{C}N} - T_{\textrm{C}1},\qquad \Theta = \frac{T_{\textrm{C}1} + T_{\textrm{C}N}}{2},\qquad T_{\textrm{C}i} = T_{\textrm{C}N} - \frac{N-i}{N-1} \Delta T.
\end{equation}
It is obvious that the accuracy of this mathematical model increases with increasing parameter $N$.

The Gaussian distribution of the local Curie temperatures $\varphi(T_{\textrm{C}i})$ in the homogeneous unit cells within the diffusion region $\Delta T$ must satisfy the boundary and normalization conditions:
\begin{equation}\label{cond}
\lim_{T_{\textrm{C}i} \to T_{\textrm{C}1} + 0} \varphi(T_{\textrm{C}i}) = \lim_{T_{\textrm{C}i} \to T_{\textrm{C}N} - 0} \varphi(T_{\textrm{C}i}) = 0,\qquad \sum_{i=1}^{N}\varphi(T_{\textrm{C}i}) = 1.
\end{equation}
Given formula~\eqref{Temp} and conditions~\eqref{cond}, one can rewrite the relation for $\varphi(T_{\textrm{C}i})$ as
\begin{equation}\label{phi}
\varphi(T_{\textrm{C}i}) = \frac{1}{\omega_{N}}\left\{\exp\left[-\frac{1}{2\eta_{N}}\left(\frac{1}{2}-\frac{N-i}{N-1}\right)^2 \right] - \exp\left(-\frac{1}{8\eta_{N}}\right)\right\},
\end{equation}
where
\begin{equation}\label{eta}
\eta_{N} = \frac{1}{N-1}\sum_{i=1}^{N}\left(\frac{1}{2}-\frac{N-i}{N-1}\right)^2,
\end{equation}
\begin{equation}\label{omega}
\omega_{N} = \sum_{i=1}^{N}\exp\left[-\frac{1}{2\eta_{N}}\left(\frac{1}{2}-\frac{N-i}{N-1}\right)^2\right] - N\exp\left(-\frac{1}{8\eta_{N}}\right)
\end{equation}
are constants which are fixed for a given $N$ value.

Taking eguations~\eqref{Ps} and \eqref{Temp} into account, one can find the temperature dependence of optical rotation $\rho_{i}$ for a given homogeneous cell with the local Curie temperature $T_{\textrm{C}i}$:
\begin{equation}\label{rhoi}
\rho_{i} =
\begin{cases}
A[T_{\textrm{C}N} - \Delta T (N-i) / (N-1) - T]^\beta, \quad T < T_{\textrm{C}i},\\
0, \quad T \geqslant T_{\textrm{C}i},
\end{cases}
\end{equation}
where the coefficient of proportionality $A$ is assumed to be the same for all homogeneous cells of the sample under study.

Considering equations~\eqref{phi} and \eqref{rhoi}, we arrive at the final temperature dependence of the optical rotation $\rho$, which is valid for the whole sample:
\begin{equation}\label{rho}
\rho = \sum_{i=1}^{N}\rho_{i}\varphi(T_{\textrm{C}i}).
\end{equation}
It follows from equation~\eqref{rho} that the equality $\rho = 0$ holds true at $T \geqslant T_{\textrm{C}N}$. In other words, the parameter $T_{\textrm{C}N}$ is the temperature at which the optical rotation $\rho$ induced by spontaneous polarization vanishes completely in the process of heating of the sample. The parameter $T_{\textrm{C}N}$ coincides with the Curie temperature found by the method I [i.e., $T_{\textrm{C}N} \equiv T_{\textrm{C}}(\textrm{I})]$. Therefore, $T_{\textrm{C}N}$ can be found directly from the experimental dependence $\rho = \rho(T)$.

Summing up, our technique for determining the CEOP at the diffuse PT implies fitting the temperature dependence of the optical rotation $\rho$ to relation~\eqref{rho} and finding the constants $T_{\textrm{C}N}$, $N$, $\eta_{N}$ and $\omega_{N}$. This enables one to determine the parameters $A$, $\Delta T$ and $\beta$, which provide the best agreement of the fitting curve with the experimental dependence. The appropriate goodness of fit is characterized by the determination coefficient $R^2$.

\section{Results and discussion}

\begin{figure}[htb]
\centerline{\includegraphics[width=0.45\textwidth]{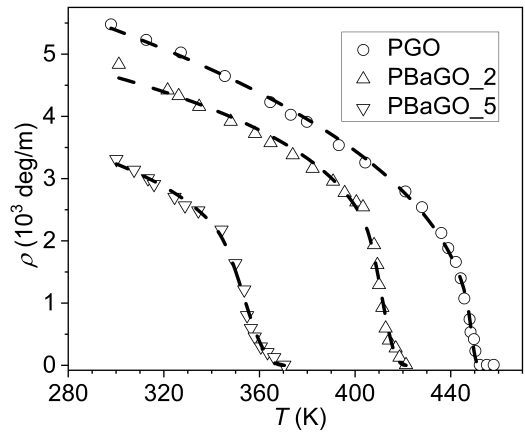}}
\caption{Temperature dependences of optical rotation $\rho$ for the pure PGO crystals and the PBaGO solid solutions ($\lambda = 632.8$~nm)~\cite{Vlkh77,Vlkh81}. Open points refer to experimental results and dashed curves refer to fitting by equation~\eqref{rho}.}
\label{f-PBaGO}
\end{figure}

\begin{figure}[htb]
\centerline{\includegraphics[width=0.45\textwidth]{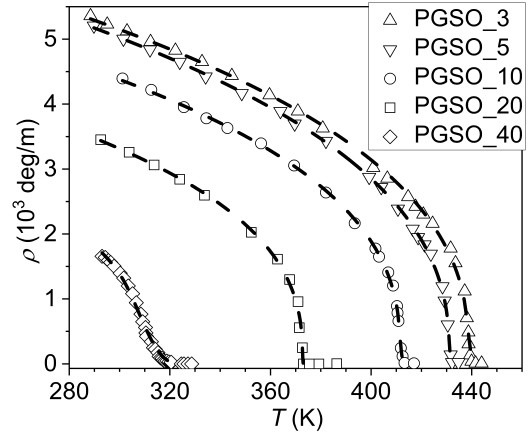}}
\caption{Temperature dependences of optical rotation $\rho$ for the PGSO solid solutions ($\lambda = 632.8$~nm)~\cite{Vlkh77,Vlkh81}. Open points refer to experimental results and dashed curves refer to fitting by equation~\eqref{rho}.}
\label{f-PGSO}
\end{figure}

Let us consider temperature dependences of the optical rotation $\rho$ obtained in works~\cite{Vlkh77,Vlkh81} for PBaGO solid solutions with the concentrations of Ba\textsuperscript{2+} ions equal to 0.02 (PBaGO\_2) and 0.05 (PBaGO\_5), as well as PGSO solid solutions with the concentrations 0.03 (PGSO\_3), 0.05 (PGSO\_5), 0.10 (PGSO\_10), 0.20 (PGSO\_20) and 0.40 (PGSO\_40) of Si\textsuperscript{4+} ions. The appropriate results are displayed in figure~\ref{f-PBaGO} and figure~\ref{f-PGSO}, respectively.

As seen from figure~\ref{f-PBaGO} and figure~\ref{f-PGSO}, the fitting curves agree well with the temperature dependences of optical rotation $\rho$ for the both PBaGO and PGSO solid solutions. The corresponding fitting parameters are listed in table~\ref{t-PBaGO} and table~\ref{t-PGSO}, respectively. For comparison, these tables also show the $T_{\textrm{C}}$ and $\beta$ values determined by the methods I [$T_{\textrm{C}}(\textrm{I})$ and $\beta(\textrm{I})$] and II [$T_{\textrm{C}}(\textrm{II})$ and $\beta(\textrm{II})$], as well as the differences $\Delta T_{\textrm{C}} = T_{\textrm{C}}(\textrm{I}) - T_{\textrm{C}}(\textrm{II})$.
\begin{table}[htb]
\caption{Main fitting parameters of temperature dependences of the optical rotation $\rho$ found for the pure PGO crystals and for the PBaGO solid solutions.}
\label{t-PBaGO}
\renewcommand{\arraystretch}{1.25}
\begin{center}
\small
\begin{tabular} {|l||r|r|r|}
\hline
 & PGO & PBaGO\_2 & PBaGO\_5\\
\hline
\hline
$\Theta$, K & 449.4 & 412.6 & 356.1\\
\hline
$T_{\textrm{C}}(\textrm{I})$,~K & 452.1 & 421.3 & 370.4\\
\hline
$T_{\textrm{C}}(\textrm{II})$,~K~\cite{Kush08} & 447.0 & 403.0 & 341.0\\
\hline
$\Delta T$,~K & $5.47±0.42$ & $17.46±0.86$ & $28.62±2.08$\\
\hline
$\Delta T_{\textrm{C}}$,~K & 5.1 & 18.3 & 29.4\\
\hline
$\beta$ & $0.40±0.01$ & $0.27±0.02$ & $0.27±0.04$\\
\hline
$\beta(\textrm{I})$ & 0.45 & 0.39 & 0.48\\
\hline
$\beta(\textrm{II})$ & 0.42 & 0.30 & 0.26\\
\hline
$R^2$ & 0.998 & 0.994 & 0.994\\
\hline
\end{tabular}
\end{center}
\end{table}
\begin{table}[!htb]
\caption{Main fitting parameters of temperature dependences of the optical rotation $\rho$ found for the PGSO solid solutions.}
\label{t-PGSO}
\renewcommand{\arraystretch}{1.25}
\begin{center}
\small
\begin{tabular} {|l||r|r|r|r|r|}
\hline
 & PGSO\_3 & PGSO\_5 & PGSO\_10 & PGSO\_20 & PGSO\_40\\
\hline
\hline
$\Theta$, K & 439.1 & 431.3 & 411.8 & 372.8 & 311.6\\
\hline
$T_{\textrm{C}}(\textrm{I})$,~K & 441.5 & 432.2 & 413.3 & 373.1 & 324.5\\
\hline
$T_{\textrm{C}}(\textrm{II})$,~K~\cite{Kush08} & 438.7 & 429.4 & 410.5 & 371.0 & 310.3\\
\hline
$\Delta T$,~K & $4.72±0.27$ & $1.82±0.17$ & $2.95±0.14$ & $0.62±0.15$ & $25.80±3.98$\\
\hline
$\Delta T_{\textrm{C}}$,~K & 2.8 & 2.8 & 2.8 & 2.1 & 14.2\\
\hline
$\beta$ & $0.39±0.01$ & $0.39±0.01$ & $0.37±0.01$ & $0.37±0.01$ & $0.40±0.16$\\
\hline
$\beta(\textrm{I})$ & 0.42 & 0.41 & 0.40 & 0.39 & --~*\\
\hline
$\beta(\textrm{II})$ & 0.40 & 0.38 & 0.36 & 0.37 & --~*\\
\hline
$R^2$ & 0.998 & 0.999 & 0.999 & 0.996 & 0.994\\
\hline
\multicolumn{6}{|l|}{* Correct calculation is not possible.}\\
\hline
\end{tabular}
\end{center}
\end{table}

Note that the fitting results obtained for pure PGO indicate that the PT in this crystal is also diffuse (see figure~\ref{f-PBaGO} and table~\ref{t-PBaGO}). This may be the main reason why the $T_{\textrm{C}}$ and $\beta$ values obtained in different works for the pure PGO crystals are different.

\begin{figure}[htb]
\centerline{\includegraphics[width=0.45\textwidth]{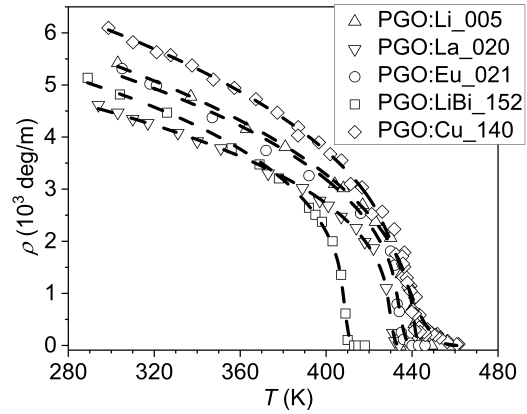}}
\caption{Temperature dependences of optical rotation $\rho$ for the doped PGO:Li\_005, PGO:La\_020, PGO:Eu\_021, PGO:LiBi\_152 and PGO:Cu\_140 crystals ($\lambda = 632.8$~nm)~\cite{Adam09,Shop09}. Open points refer to experimental results and dashed curves refer to fitting by equation~\eqref{rho}.}
\label{f-PGO:D}
\end{figure}

Now, let us analyze the temperature dependences of the optical rotation $\rho$ obtained in our previous works~\cite{Adam09,Shop09} for the PGO crystals doped with 0.005 wt.~\% of Li\textsuperscript{3+} (PGO:Li\_005), 0.020 wt.~\% of La\textsuperscript{3+} (PGO:La\_020), 0.021 wt.~\% of Eu\textsuperscript{3+} (PGO:Eu\_021), 0.152 wt.~\% of Li\textsuperscript{3+} and Bi\textsuperscript{3+} (PGO:LiBi\_152) and 0.14 wt.~\% of Cu\textsuperscript{2+} (PGO:Cu\_140) (see figure~\ref{f-PGO:D}). Using our technique for determining the CEOPs at the diffuse PTs, one can find the fitting parameters listed in table~\ref{t-PGO:D}.

It is worth noting that the fitting parameters presented in tables~\ref{t-PBaGO}, \ref{t-PGSO} and \ref{t-PGO:D} are obtained for the case $N = 10^{3}$. Our studies have also testified that further increasing $N$ value does not lead to a significant increase in the accuracy of the model and to a better correspondence of experiment and theory. As seen from tables~\ref{t-PBaGO}, \ref{t-PGSO} and \ref{t-PGO:D}, the inequality $T_{\textrm{C}}(\textrm{II})\leqslant\Theta\leqslant T_{\textrm{C}}(\textrm{I})$ is commonly valid for the $\Theta$ and $T_{\textrm{C}}$ parameters. In its turn, the difference $\Delta T_{\textrm{C}}$ is comparable with the fitting parameter $\Delta T$. Therefore, the methods I and~II combined together can be considered as a rapid test for estimating a degree of diffusion of the PT.

\begin{table}[!htb]
	\caption{Main fitting parameters of temperature dependences of the optical rotation $\rho$ found for the doped crystals PGO:Li\_005, PGO:La\_020, PGO:Eu\_021, PGO:LiBi\_152 and PGO:Cu\_140.}
	\label{t-PGO:D}
	\renewcommand{\arraystretch}{1.25}
	\begin{center}
		\small
		\begin{tabular} {|l||r|r|r|r|r|}
			\hline
			& PGO:Li\_005 & PGO:La\_020 & PGO:Eu\_021 & PGO:LiBi\_152 & PGO:Cu\_140\\
			\hline
			\hline
			$\Theta$, K & 441.4 & 430.5 & 435.0 & 409.3 & 440.6\\
			\hline
			$T_{\textrm{C}}(\textrm{I})$,~K & 444 & 434 & 440 & 413 & 460.8\\
			\hline
			$T_{\textrm{C}}(\textrm{II})$,~K & 440 & 428 & 432 & 408 & 432.0\\
			\hline
			$\Delta T$,~K & $5.19±0.38$ & $7.07±0.35$ & $9.95±1.34$ & $7.46±0.48$ & $40.42±1.93$\\
			\hline
			$\Delta T_{\textrm{C}}$,~K & 4 & 6 & 8 & 5 & 28.8\\
			\hline
			$\beta$ & $0.41±0.01$ & $0.34±0.01$ & $0.37±0.03$ & $0.33±0.01$ & $0.38±0.02$\\
			\hline
			$\beta(\textrm{I})$ & 0.42 & 0.37 & 0.48 & 0.42 & 0.47\\
			\hline
			$\beta(\textrm{II})$ & 0.39 & 0.33 & 0.43 & 0.39 & 0.34\\
			\hline
			$R^2$ & 0.998 & 0.999 & 0.991 & 0.997 & 0.990\\
			\hline
		\end{tabular}
	\end{center}
\end{table}

Given the above results, one can conclude that the method for determining the CEOPs at diffuse PTs, which is used in the present work, has notable advantages over the other methods employed for finding the Curie temperature $T_{\textrm{C}}$ and determining the CEOP $\beta$. For instance, the choice of $T_{\textrm{C}}$ as a point where the optical rotation $\rho$ vanishes completely in the course of heating of a sample (the method I) is not correct, provided that a ``tail'' of optical rotation is observed at the DFEPT. The definition of $T_{\textrm{C}}$ as a point of minimum of the temperature dependence of the derivative $\rd \rho^{2} / \rd T$ (the method II)~\cite{Kush08} is also not indisputable. Indeed, in case of $\beta < 0.5$, given equation~\eqref{Ps}, the parameter $\rd \rho^{2} / \rd T \propto -2\beta/(T_{\textrm{C}} - T)^{1-2\beta}$ at $T = T_{\textrm{C}}$ should tend to $-\infty$ rather than to a finite value (see figure~\ref{smthCu}).

The other point is that determination of the CEOP $\beta$ requires analyzing an additional dependence, $\log \rho$ vs. $\log (T_{\textrm{C}} - T)$. Moreover, another temperature dependence, that of the derivative $\rd \rho^{2} / \rd T$, must be plotted in order to determine $T_{\textrm{C}}$ in case of the method II. More important, selection of that part of logarithmic dependence which should be fitted by a linear function is rather subjective. On the contrary, the approach to the calculation of CEOP presented in this work relies only upon the fitting procedure for the dependence $\rho = \rho(T)$ and the $T_{\textrm{C}N}$ data which can be found using simple and standard objective techniques for interpolation of the experimental temperature dependence of optical activity. Finally, our approach enables one to determine the diffusion region $\Delta T$ for the PT, in contrast to the other methods.

\section{Conclusions}

In the present work, we have described the method suggested for accurate determination of the CEOPs in the PGO-based solid solutions that manifest the DFEPTs. This method consists in dividing a crystal sample under study into a large number of homogeneous unit cells, each of which has a non-diffuse PT with exactly defined local Curie temperature. Then, we fit the temperature dependences of the optical rotation, which are proportional to the spontaneous polarization (i.e., the order parameter), using a straightforward phenomenological relation. As a result, we are able to find the CEOP itself, the region where the PT is diffuse, and the average Curie temperature.

Using this method, we have determined the CEOPs for the pure Pb\textsubscript{5}Ge\textsubscript{3}O\textsubscript{11} crystals, the solid solutions Pb\textsubscript{5}(Ge$_{1-x}$Si$_{x}$)\textsubscript{3}O\textsubscript{11} ($x = 0.03$, 0.05, 0.10, 0.20, 0.40) and (Pb$_{1-x}$Ba$_{x}$)\textsubscript{5}Ge\textsubscript{3}O\textsubscript{11} ($x = 0.02$, 0.05), and the doped crystals Pb\textsubscript{5}Ge\textsubscript{3}O\textsubscript{11}:Li\textsuperscript{3+} (0.005 wt.~\%), Pb\textsubscript{5}Ge\textsubscript{3}O\textsubscript{11}:La\textsuperscript{3+} (0.02 wt.~\%), Pb\textsubscript{5}Ge\textsubscript{3}O\textsubscript{11}:Eu\textsuperscript{3+} (0.021~wt.~\%), Pb\textsubscript{5}Ge\textsubscript{3}O\textsubscript{11}:Li\textsuperscript{3+}, Bi\textsuperscript{3+} (0.152 wt.~\%) and Pb\textsubscript{5}Ge\textsubscript{3}O\textsubscript{11}:Cu\textsuperscript{2+} (0.14 wt.~\%). Our method is compared with the other approaches known from the literature, which also determine the parameters of~PT. Significant advantages of our approach are highlighted.

\ukrainianpart

\title{Критичні показники параметра порядку розмитих сегнетоелектричних фазових переходів у твердих розчинах на основі германату свинцю: дослідження оптичної активності}
\author{Д. І. Адаменко, Р. О. Влох}
\address{Iнститут фізичної оптики iменi О. Г. Влоха, вул. Драгоманова, 23, місто Львів, 79005, Україна}

\makeukrtitle

\begin{abstract}
\tolerance=10000
У цій роботі показано, що критичні показники параметра порядку розмитих сегнетоелектричних фазових переходів, що відбуваються в кристалах сімейства германату свинцю, можна визначити за допомогою експериментальних температурних залежностей їх оптичної активності. Також описано підхід, який передбачає поділ кристалічного зразка на безліч однорідних елементарних комірок, кожна з яких харак\-теризується нерозмитим фазовим переходом із певною локальною температурою Кюрі. Використову\-ючи цей підхід, було визначено критичні показники параметра порядку для кристалів чистого Pb\textsubscript{5}Ge\textsubscript{3}O\textsubscript{11}, твердих розчинів Pb\textsubscript{5}(Ge$_{1-x}$Si$_{x}$)\textsubscript{3}O\textsubscript{11} ($x =$~0.03, 0.05, 0.10, 0.20, 0.40) і (Pb$_{1-x}$Ba$_{x}$)\textsubscript{5}Ge\textsubscript{3}O\textsubscript{11} ($x =$~0.02, 0.05) та легованих кристалів Pb\textsubscript{5}Ge\textsubscript{3}O\textsubscript{11}:Li\textsuperscript{3+} (0.005~ваг.~\%), Pb\textsubscript{5}Ge\textsubscript{3}O\textsubscript{11}:La\textsuperscript{3+} (0.02 ваг.~\%), Pb\textsubscript{5}Ge\textsubscript{3}O\textsubscript{11}:Eu\textsuperscript{3+} (0.021 ваг.~\%), Pb\textsubscript{5}Ge\textsubscript{3}O\textsubscript{11}:Li\textsuperscript{3+}, Bi\textsuperscript{3+} (0.152 ваг.~\%) і Pb\textsubscript{5}Ge\textsubscript{3}O\textsubscript{11}:Cu\textsuperscript{2+} (0.14 ваг.~\%). Порівняння даного підходу з іншими методиками визначення температур Кюрі та критичних показників параметра порядку розмитих сегнетоелектричних фазових переходів свідчить про його істотні переваги.
\keywords тверді розчини, сегнетоелектричний фазовий перехід, критичні показники, оптична активність

\end{abstract}

\lastpage
\end{document}